\documentclass{sistedes}

\hypersetup{
    colorlinks,
    linkcolor={black!50!black},
    citecolor={blue!50!black},
    urlcolor={blue!80!black}
}
\usepackage{graphicx}

\begin{document}
\title{QuantumX: an experience for the consolidation of Quantum Computing and Quantum Software Engineering as an emerging discipline}
\titlerunning{QuantumX}
%
\author{Juan M. Murillo\inst{1}\orcidID{0000-0003-4961-4030} \and
Ignacio García Rodríguez de Guzmán\inst{2}\orcidID{0000-0002-0038-0942} \and
Enrique Moguel\inst{1}\orcidID{0000-0002-4096-1282} \and
Javier Romero-Alvarez\inst{1} \and
Jaime Alvarado-Valiente\inst{1} \and
Álvaro M. Aparicio-Morales\inst{1} \and
Jose García-Alonso\inst{1} \and
Ana Díaz Muñoz\inst{3} \and
Eduardo Fernández-Medina\inst{2} \and
Francisco Chicano\inst{4} \and
Carlos Canal\inst{4} \and
José Daniel Viqueira\inst{5}\inst{6} \and
Sebastián Villarroya\inst{6} \and
Eduardo Gutiérrez\inst{6} \and
Adrián Romero-Flores\inst{7} \and
Alfonso E. Márquez-Chamorro\inst{7} \and
Antonio Ruiz-Cortes\inst{7} \and
Cyrille YetuYetu Kesiku\inst{8} \and
Pedro Sánchez\inst{9} \and
Diego Alonso Cáceres\inst{9} \and
Lidia Sánchez-González\inst{10}\orcidID{0000-0002-0760-1170} \and
Fernando Plou\inst{11} \and
Ignacio Aguilar\inst{12}
}
\authorrunning{Murillo and García Rodríguez de Guzmán et al.}
%
\institute{
Quercus Software Engineering Group, Universidad de Extremadura\\
\email{juanmamu@unex.es, enrique@unex.es}\\
\and
ITSI, Universidad de Castilla-La Mancha\\
\email{Ignacio.GRodriguez@uclm.es}\\
\and
AQCLab Software Quality / University of Castilla-La Mancha\\
\and
ITIS Software, Universidad de Málaga\\
\and
Galicia Supercomputing Center (CESGA)\\
\and 
COGRADE, Universidade de Santiago de Compostela\\
\and
SCORE Lab, I3US Institute, Universidad de Sevilla\\
\and
University of Deusto\\
\and
Universidad Politécnica de Cartagena\\
\and
Robotics Group, I4 Institute, Universidad de León\\
\and
University of Oviedo\\
\and
PAPC, Universidad de Málaga\\
}

\maketitle              

\begin{abstract}
The first edition of the QuantumX track, held within the XXIX Jornadas de Ingeniería del Software y Bases de Datos (JISBD 2025), brought together leading Spanish research groups working at the intersection of Quantum Computing and Software Engineering. The event served as a pioneering forum to explore how principles of software quality, governance, testing, orchestration, and abstraction can be adapted to the quantum paradigm. The presented works spanned diverse areas (from quantum service engineering and hybrid architectures to quality models, circuit optimization, and quantum machine learning), reflecting the interdisciplinary nature and growing maturity of Quantum Computing and Quantum Software Engineering. The track also fostered community building and collaboration through the presentation of national and Ibero-American research networks such as RIPAISC and QSpain, and through dedicated networking sessions that encouraged joint initiatives. Beyond reporting on the event, this article provides a structured synthesis of the contributions presented at QuantumX, identifies common research themes and engineering concerns, and outlines a set of open challenges and future directions for the advancement of Quantum Software Engineering. This first QuantumX track established the foundation for a sustained research community and positioned Spain as an emerging contributor to the European and global quantum software ecosystem.

\end{abstract}
%
%
%

\section{Introduction}

The track “\textit{Quantum Computing and Quantum Software Engineering}” (QuantumX), held on September 10 and 11, 2025, as part of the 29th Conference on Software Engineering and Databases (JISBD 2025)\footnote[1]{\url{https://www.uco.es/sistedes2025/ProgramaJISBD_QUANTUM.php}}, organized by SISTEDES and the University of Córdoba, was the first national forum specifically dedicated to the convergence between quantum computing and software engineering.

This pioneering track, in its first edition, brought together a diverse group of established and emerging research groups from universities and institutions across the country to explore the challenges, methodologies, and opportunities that arise when applying software engineering principles to the development of quantum solutions, as well as when designing tools that facilitate the adoption of this disruptive technology.

Quantum computing is currently in a transitional stage between theory and practice, marked by the availability of restricted-access hardware, rapid evolution of programming languages, and the need to establish good engineering practices that ensure the quality, maintainability, and reliability of quantum software \cite{Piattini2020}. In this context, interdisciplinary collaboration between software engineers, physicists, mathematicians, and artificial intelligence experts is essential to overcome the barriers that still limit the industrial application of quantum technology \cite{Murillo2025}.

The QuantumX track has therefore served as a meeting and collaboration space for the main Spanish groups leading research in quantum computing and quantum software engineering, with the common goal of creating a cohesive community and an international reference point. Throughout the sessions, papers were presented addressing aspects such as quantum circuit planning and orchestration, quantum service engineering, quality metrics for hybrid quantum-classical software, quantum machine learning models, and the design of abstractions and high-level quantum programming languages, among others.

Beyond documenting the event, this article aims to provide a structured overview and critical synthesis of the research contributions presented at the QuantumX track. By analyzing the topics, approaches, and engineering concerns addressed by the participating works, the paper identifies common research themes, emerging trends, and open challenges that characterize the current state of Quantum Software Engineering. In doing so, it seeks to contribute to the consolidation of this field by offering a reference point for researchers and practitioners interested in the systematic application of software engineering principles to quantum computing.

To illustrate the breadth and diversity of research represented at the track, the following participating groups are highlighted:

\begin{itemize}
    \item \textbf{Quercus Software Engineering Group} (University of Extremadura), a well-established group in Software Engineering with a notable line of research dedicated to the development of methodologies, tools, and quality models for quantum software.
    
    \item \textbf{Alarcos} (University of Castilla-La Mancha), a national leader in software quality, assurance, and hybrid software governance, with recent contributions to quality models, metrics, and evaluation environments for quantum and hybrid systems.

    \item \textbf{ITIS Software} (University of Malaga) works on Software Engineering, focusing on software architecture, quality, and high-level abstractions for quantum software, as well as topics at the intersection of quantum computing, artificial intelligence, and post-quantum cryptography.

    \item \textbf{Parallel Architectures, Programming, and Compilers} (PAPC, University of Malaga) focuses on parallel programming and high-performance architectures, with recent work on quantum and hybrid quantum-classical computing, including quantum machine learning and performance analysis.
    
    \item \textbf{Quantum and High Performance Computing} (QHPC, University of Oviedo and CSIC), focused on the intersection between high-performance computing and quantum machine learning, with a strong emphasis on experimentation and tool development.
    
    \item \textbf{Computer Graphics and Data Engineering} (COGRADE, University of Santiago de Compostela), a group with a solid track record in databases, visualization, and applied computing, currently expanding its lines of research into quantum databases.
    
    \item \textbf{ROBÓTICA} (University of León), dedicated to research in robotics and computer vision, with an interest in quantum image processing and its potential in intelligent automation environments. They also work on quantum-resistant encryption and digital notary systems, QKD support, and a quantum key distribution center.
    
    \item \textbf{Engineering, Vision, Data \& Artificial Intelligence} (eVIDA, University of Deusto), internationally recognized for its work in artificial intelligence applied to health, biomedical signal processing, and, more recently, in the development of hybrid quantum-classical models for the classification and analysis of medical information.
    
    \item \textbf{DeustoTech} and \textbf{Deusto for Knowledge} (University of Deusto), focused on quantum computing applied to machine learning and the optimization of variational models.

    \item \textbf{SCORE Lab} (University of Seville) The Smart Computer Systems Research and Engineering (SCORE) lab conducts research in Software and Service Engineering, with a focus on the lifecycle management of services in complex ecosystems, particularly in hybrid quantum-classical architectures.

\end{itemize}

The participation of these groups has highlighted the richness and complementarity of the Spanish scientific ecosystem in the quantum field, where lines of research in computational architectures, parallel programming, software engineering, artificial intelligence, computer vision, and databases converge.

In addition, it is worth highlighting the collaborative and inter-university nature of many of the papers presented, in which different teams join forces and knowledge to tackle complex challenges, such as the execution of quantum mutants through advanced planning, the integration of quantum computing into hybrid solutions, or the creation of abstractions that simplify next-generation quantum programming.

The QuantumX track has shown that Spain has a solid, multidisciplinary research base in the field of quantum software engineering, with groups that not only conduct research from a theoretical perspective but also develop prototypes, tools, and technology transfer projects. This diversity of approaches is a strategic asset for the future development of the national quantum ecosystem, aligned with European priorities for digital innovation, technological sovereignty, and training of specialized talent.

\section{Research groups and papers presented}

This section introduces the research groups that participated in the track and summarizes the papers presented.

\subsection{Quercus Software Engineering Group}

Quercus Software Engineering Group at the University of Extremadura (UEx) is a well-established research group that has been active since 1996, with a continuous track record in European, national, and regional projects. Its lines of research cover Software Engineering, Artificial Intelligence, IoT, and, in recent years, Quantum Software Engineering. In addition, it actively participates in various research networks such as the Science and Engineering Services Network and digital innovation hubs.

Among the papers presented during the conference, the following stand out:

\begin{itemize}
    \item \textit{Circuit scheduling policies on current QPUs: QCRAFT Scheduler} \cite{Alvarado-Valiente2025}. This work defines and validates three planning policies for the QCRAFT Scheduler, with the aim of grouping and combining circuits in cloud providers' QPUs, reducing waiting times and costs without significantly degrading results. The authors report average reductions in cost ($\approx84$\%) and tasks ($\approx84$\%) compared to individual executions. 

    \item \textit{ATHENA-HARMONIA: Integración de la Computación Cuántica en Soluciones Híbridas para la Nueva Generación de Software} \cite{Romero-Alvarez2025}. It presents the HARMONIA subproject (part of the ATHENA project), which seeks to integrate quantum and classical components under the computational continuum paradigm (cloud-edge-quantum). It pursues engineering methodologies and tools that enable the design, deployment, and governance of hybrid solutions with the same level of abstraction as classical software.

    \item \textit{Service Engineering for Quantum Computing: Ensuring High-Quality Quantum Services} \cite{Diaz-Munoz2025b}. Work in collaboration with the Alarcos Group at the UCLM, where a quantum service engineering process is introduced to ensure quality (e.g., analyzability) through extended OpenAPI, automatic code generation, and analysis with SonarQube. It evaluates 40 implementations of quantum algorithms, highlighting variability in quality attributes and offering practical guidance for continuous improvement.

    \item \textit{Ejecutando mutantes cuánticos utilizando técnicas de planificación} \cite{Casco-Seco2025}. Work in collaboration with the Alarcos Group at the UCLM, where a tool is presented that creates and executes quantum circuit mutation tests on real hardware through task planning and combination, achieving cost reductions of up to $\approx94$\% in scenarios with IBM Quantum. It demonstrates the feasibility of large-scale quantum testing in the NISQ era.

    \item \textit{Quantum integer: Towards a higher abstraction level in quantum programming} \cite{SanchezRivero2025}. Collaboration with the SCORE Lab Group at the University of Seville, which proposes the “quantum integer” data type and a set of composable oracles (e.g., Less-Than, Greater-Than, intervals, and multiples) to raise the level of abstraction in quantum programming beyond gates and circuits, promoting reuse, maintainability, and modularity in quantum software development.

    \item \textit{Locus: una abstracción para facilitar la programación basada en circuitos cuánticos} \cite{ZayasGallardo2025}. Work in collaboration with the ITIS Group at the UMA, where a slight abstraction (“Locus”) is proposed to develop programs based on quantum circuits with better reuse and maintainability attributes, emulating the historic leap from low-level languages to high-level types/operations in classical software. It is a first step towards DSLs and reusable quantum component libraries.
\end{itemize}

\subsection{Alarcos}

Alarcos is the Software Engineering research group at the University of Castilla-La Mancha (UCLM). It is a nationally and internationally renowned team in software quality, measurement and governance, processes and standards, requirements, modeling, and empirical software engineering. In recent years, it has expanded its activity to the field of Quantum Software Engineering, where it contributes its experience in quality models and metrics, assurance and testing (including quantum mutation), and governance of hybrid quantum-classical services.

Among the papers presented during the conference, the following stand out:

\begin{itemize}
    \item \textit{Validación empírica preliminar de un modelo de analizabilidad para software híbrido} \cite{Diaz-Munoz2025a}. This paper presents a controlled validation of a quality model focused on analyzability (ISO/IEC 25010) for hybrid systems that integrate classical and quantum components. The model combines classical metrics with metrics specific to quantum circuits (e.g., depth, complexity). The results support the usefulness of the approach and lay the groundwork for further empirical studies to consolidate its validity in academic and industrial settings.

    \item \textit{Generación de circuitos de Grover a partir de matrices} \cite{Polo2025}. This paper proposes an algorithm for synthesizing Grover oracles from binary matrices and integrates it into the quco (Quantum Code Generation) tool. The paper demonstrates how to automate circuit generation and reduce low-level design friction, facilitating reuse and traceability from higher-level specifications.

    \item \textit{Minimizing incident response time in real-world scenarios using quantum computing} \cite{Serrano2025}. Work in collaboration with the GSyA group at the University of Castilla-La Mancha, presenting a quantum formulation of the incident response problem in Information Security Management Systems (ISMS). The approach prioritizes and selects mitigation controls to shorten system restoration time in the event of multiple incidents. The work provides experimental evidence in realistic scenarios and outlines a roadmap for bringing quantum optimization to managed cybersecurity services.

    \item \textit{Service Engineering for Quantum Computing: Ensuring High-Quality Quantum Services} \cite{Diaz-Munoz2025b}. Collaborative work with the Quercus Group at the University of Extremadura, proposing a quantum service engineering process that extends OpenAPI, incorporates static analysis with SonarQube, and generates code (e.g., for IBM Quantum) to automate deployment and evaluation tasks. After analyzing dozens of quantum algorithm implementations, the work identifies variability in quality attributes (e.g., analyzability) and offers practical guidelines for the continuous improvement of quantum software.

    \item \textit{Ejecutando mutantes cuánticos utilizando técnicas de planificación} \cite{Casco-Seco2025}. Work in collaboration with the Quercus Group at the University of Extremadura, introducing a quantum mutation testing tool that generates, composes, and plans the execution of circuit mutants in real hardware, combining tasks to leverage QPUs in the cloud. It demonstrates substantial cost and time reductions compared to individual executions, and provides case studies on IBM Quantum that demonstrate the viability of quantum testing in the NISQ era.
\end{itemize}

\subsection{SCORE Lab}

The SCORE Lab is the Software and Services research laboratory at the University of Seville, affiliated with the Institute for Research in Computer Engineering (I3US). Its activity focuses on the design, governance, and operation of software services in complex ecosystems (including hybrid quantum-classical services), with special emphasis on architectures, orchestration, quality, automation, smart contracts (i-contracts), and capacity and demand models. The group combines fundamental and applied research, generating methods, conceptual frameworks, and tools that facilitate the portability, efficiency, governance, and self-adaptation of distributed services. In the quantum field, SCORE Lab transfers these principles to the development of HSaaS (Hybrid Software-as-a-Service) and multi-provider orchestration in the NISQ era.

Among the papers presented during the conference, the following stand out:

\begin{itemize}
    \item \textit{Qrchestrator: An Error and Load Aware Quantum Orchestrator for Multiple NISQ Cloud Providers} \cite{Romero-Flores2025}. This paper presents a quantum orchestrator designed to optimize circuit execution across different NISQ providers, considering both workload and hardware error and fidelity rates. This approach allows for dynamic selection of the most suitable backend, reducing failed executions and improving overall efficiency. In doing so, the work lays the foundation for multi-cloud quantum orchestration engineering, aligned with the reliability and automation practices typical of classic distributed services.
    
    \item \textit{An Initial Exploration of Pricing-driven Governance for Hybrid Quantum-Classical SaaS} \cite{Ruiz-Cortes2025}. This work proposes an initial model of price-based governance for hybrid quantum-classical services (HSaaS). Through smart contracts and self-adaptive policies, the system dynamically adjusts its configuration based on cost, load, and demand, balancing performance and economic sustainability. This approach introduces an innovative vision of automated governance in the quantum context, combining technical and financial metrics in service management.

    \item \textit{Quantum integer: Towards a higher abstraction level in quantum programming} \cite{SanchezRivero2025}. This work has been developed in collaboration with the Quercus Group at the University of Extremadura (UEx) and introduces the concept of quantum integers as a new programming abstraction. This type of data, together with composable oracles, allows for the development of more modular and reusable quantum programs, raising the level of abstraction and reducing the complexity of circuit design. The work moves towards more structured and maintainable quantum programming, in line with the principles of classical software engineering.
\end{itemize}

\subsection{ITIS}

The ITIS Software Group at the University of Malaga (UMA) is a research team focused on software engineering, with a special emphasis on languages, methods, and tools for the development of complex, high-performance systems. Its lines of research cover software architecture and quality, programming and verification, big data and artificial intelligence, and in recent years, it has incorporated a growing focus on applied quantum computing, exploring abstractions, models, and best practices that transfer the principles of modularity, reuse, and maintainability that characterize classical software to the quantum domain. ITIS also focuses on quantum and post-quantum cryptography and in research lines at the intersection between quantum computing and artificial intelligence: quantum artificial intelligence (QAI) and AI4Quantum.

This research group presented a paper entitled “\textit{Locus: una abstracción para facilitar la programación basada en circuitos cuánticos}” \cite{ZayasGallardo2025} (in collaboration with the Quercus Group at the University of Extremadura), which proposes a lightweight abstraction for programming at a higher level on quantum circuits. The work is based on the historical analogy between “gate” programming and assembler programming, and proposes to raise the abstraction level through reusable components and higher-level operations that promote clarity, maintainability, and composition of quantum software. Instead of defining each circuit from scratch, Locus suggests a construction vocabulary that reduces design friction, improves traceability, and facilitates collaboration between teams, laying the groundwork for DSLs and libraries geared toward the practice of Quantum Software Engineering. During the presentation, the authors also mentioned the Quant·UMA initiative\footnote{\url{https://quant.uma.es}} of the University of Malaga, where the different research groups working on quantum technologies join forces to work on training programs and disseminate the research on the topic.

\subsection{PAPC}

The PAPC (Parallel Architectures, Programming, and Compilers) Group at the University of Malaga (UMA) conducts research at the intersection of parallel architectures, compilers, and high-performance programming, with a special focus on the optimization, performance, and scalability of scientific and machine learning software. In recent years, the group has also developed a growing and ambitious research line in quantum technologies and quantum machine learning, applying its expertise in performance engineering and parallel computing to accelerate and improve the reproducibility of hybrid quantum-classical workflows, enable large-scale simulation of quantum computation using HPC resources, and develop new quantum algorithms for the NISQ era. This group is also actively involved in the Quant·UMA initiative.

The research group presents the work entitled “\textit{Redes tensoriales aplicadas a redes neuronales cuánticas: evaluación de eficiencia y rendimiento}” \cite{Aguilar2025}, which provides a systematic evaluation of tensor networks (TNs), as feature extractors placed before the variational quantum circuit (VQC) within a hybrid quantum neural network (HQNN) architecture. The study analyzes how TNs can improve the alignment of input data with the quantum block through a quantum-inspired processing scheme. Early results suggest that, for specific architectures and tasks, TNs can transform data using fewer parameters and prepare it more effectively for the quantum circuit, while also highlighting limitations and best practices—such as the choice of TN architecture, bond-dimension control, and variational circuit design—to maintain predictive performance at a reasonable computational cost.

\subsection{D4K}

The DeustoTech / Deusto for Knowledge (D4K) group at the University of Deusto promotes applied research in artificial intelligence, quantum computing, and data technologies, with a special focus on transfer to domains such as health, industry, and digital services. Its approach combines model and algorithm design with software engineering and reproducible experimentation, exploring hybrid quantum-classical architectures, machine learning techniques, and tools to accelerate scientific validation and technological adoption.

D4K presented the paper entitled “\textit{Shaping Expressibility in Variational Circuits: Balancing Trainability and Performance through Experimental Analysis}” \cite{HernandezLopez2025}, in which researchers experimentally study how two mechanisms (bandwidth tuning and latent qubits) affect the expressiveness and trainability of variational quantum circuits (VQCs) used as classifiers. The results show that adjusting the bandwidth does not substantially improve trainability and can reduce expressiveness, making it unsuitable for QNNs in general. In contrast, adding latent qubits increases trainability and allows for better results in already trainable circuits, expanding the feature space with minimal loss of expressiveness (although with decreasing returns as the number of qubits increases). The work provides practical guidelines for the design of scalable VQCs and suggests combinations of techniques that balance predictive quality and training feasibility in hybrid environments.

\subsection{eVIDA}

The eVIDA (Engineering, Vision, Data \& Artificial Intelligence) group at the University of Deusto is a multidisciplinary team focused on applying AI, computer vision, and data technologies to real-world problems (especially in digital health and the biomedical field), combining methodological research with technology transfer. Its work ranges from the design of machine learning models and pipelines to clinical validation and the creation of reproducible tools, and in recent years, it has incorporated a specific line of research into quantum machine learning and quantum-classical integration for signal analysis and clinical text tasks.

Their work, entitled “\textit{Harnessing the quantum parallel model to improve clinical text classification}” \cite{YetuyteuKesiku2025} presents a hybrid model that combines a classic 1D CNN with a quantum BiLSTM and a quantum attention layer for the classification of medical texts, focusing on reports related to lung cancer. BiLSTM cells integrate variational circuits into gate functions, and attention leverages entanglement to improve context, achieving F1 and MCC improvements over classical bases in MIMIC-III/IV datasets. The work shows that quantum parallelization and variational encoding can bring measurable gains in clinical NLP, and offers a concrete, reusable architecture for advancing precision medicine and decision support systems.

\subsection{COGRADE / CESGA}
The partnership between the Galicia Supercomputing Center (CESGA) and the COGRADE group at the University of Santiago de Compostela (USC) integrates high-performance computing (HPC) with advanced research in data engineering and visual computing. CESGA provides the computational framework, including scientific storage and cloud infrastructure, necessary to scale multidisciplinary projects. Meanwhile, COGRADE provides the analytical expertise required to design algorithms that translate theoretical models into practical applications. Their collaboration in quantum computing specifically targets hybrid methodologies and efficient emulation to facilitate experimentation beyond the constraints of NISQ hardware.

CESGA presented the paper entitled “\textit{Density Matrix Emulation of Quantum Recurrent Neural Networks for Multivariate Time Series Prediction}” \cite{ViqueiraCao2025} which proposes emulation using density matrices of quantum recurrent neural networks (QRNNs) for the prediction of multivariate time series. The idea is to capture the characteristic behavior of quantum recurrent architectures in a more stable and scalable simulation scheme, maintaining essential properties (superposition, entanglement) and avoiding the practical limitations of NISQ hardware. The approach facilitates the training and evaluation of quantum models in HPC environments with greater control over noise and resources, allowing for the comparison of configurations, the study of hyperparameters, and the exploration of potential advantages over classical alternatives before making the leap to actual hardware. Overall, the work offers a pragmatic way to mature QML applied to complex time series prediction, accelerating the experimentation cycle and reducing the barrier to entry for quantum model engineering.

The paper presented by COGRADE, entitled “\textit{Diseño e implementación de una base de datos cuántica}” \cite{Gutierrez2025}, outlines a strategic roadmap for integrating quantum technologies into database management.The study focuses on two primary areas: indexing and query optimization within relational environments. It proposes exploring the application of quantum algorithms (e.g., Grover), alongside hybrid classical-quantum approaches to evaluate their impact on response times, accuracy, security, and transaction management. Furthermore, the research plan includes prototypes developed using Qiskit, tested in simulators and quantum hardware, to analyze the effect of NISQ noise and compare alternative implementations. Ultimately, the article outlines key challenges and opportunities for transferring quantum advantages to core DBMS tasks, establishing a clear experimental and technological agenda for their progressive validation.

\subsection{DSIE}

The DSIE (Division of Systems and Electronics Engineering) Research Group at the Universidad Politécnica de Cartagena (UPCT) works at the intersection of systems engineering, computer science, and electronics, with a special focus on algorithms and optimization, high-performance computing, and, more recently, applied quantum computing. Its activity combines methodological research with the development of prototypes and tools to solve combinatorial and engineering problems, collaborating with other national groups on software quality and emerging technologies.

In their paper entitled “\textit{Optimizaciones del oráculo de Grover mediante el uso de contadores para el problema del coloreado de grafos}” \cite{Alonso2025}, they present a graph coloring-specific Grover oracle construction that introduces register-based counters to compactly verify coloring constraints (e.g., conflicts between adjacent vertices). The idea is to replace part of the intensive multi-control logic with counter subcircuits and local checks, which reduces the circuit depth and the number of multi-control/auxiliary gates compared to direct approaches. The work discusses the encoding scheme, the composition of counters within the oracle, and the impact on NISQ resources (gates, depth, possible accumulated errors), showing that this strategy makes Grover search more efficient and scalable in coloring instances, and is extrapolatable to other combinatorial search/optimization tasks with similar constraints.

\subsection{CSIC / QHPC}

The IPLA Dairy Research Institute is a center of the Spanish National Research Council (CSIC) focused on biotechnology and data science applied to food and health, with experience in signal analysis, statistical modeling, and knowledge transfer to real-world problems. The Quantum and High Performance Computing (QHPC) group at the University of Oviedo focuses on high-performance computing and quantum computing, developing tools and methodologies to accelerate reproducible research in machine learning (classical and quantum) using HPC clusters and hybrid environments. Together, they form a tandem that combines use cases and experimental rigor with software engineering and performance.

The work presented by both groups, entitled “\textit{A Lightweight Python Package for Quantum Machine Learning Model Benchmarking}” \cite{Garcia-Vega2025}, introduces LazyQML, a lightweight Python package for training, comparing, and classifying Quantum Machine Learning models in supervised tasks (binary and multi-class) from a unified environment. The tool allows multiple variants of parameterizable models and circuits to be run on CPU or parallel environments, reporting comparable metrics (e.g., balanced accuracy) and facilitating informed decisions about architectures, feature maps, and training configurations. Its proposal reduces the typical friction of QML (dependencies, ad hoc scripts, and lack of comparability) by offering a reproducible battery of benchmarks that accelerates model exploration before moving on to quantum hardware or more expensive pipelines.

\subsection{ROBÓTICA}

The ROBÓTICA Group at the University of León works on social robotics, human-robot interaction, explainability, and cognitive architectures, conducting experiments in real-world environments. Through collaboration with the National Cybersecurity Institute (INCIBE), it develops cybersecurity solutions applied to cyber-physical systems. Its research lines also include other aspects related to robotics, such as simulation, computer vision, and haptic devices. In the field of quantum computing, they also work on the quantum processing of digital images and, related to cybersecurity, have collaborated on projects with SCAYLE, creating quantum-resistant encryption and digital notary systems, QKD support, and a quantum key distribution center.

In their paper entitled ``\textit{Revisión de las técnicas existentes para el procesamiento cuántico de imágenes}'' \cite{Merino2025}, they present a critical state-of-the-art review of quantum computer vision: image representation/compression, scaling and fusion, edge detection, classification with QNNs, architecture optimization, segmentation, and sectoral applications. The main conclusion is that the current literature suffers from a lack of rigor and reproducibility (poorly specified circuits, absence of open source code, tests on tiny patches, little comparison with classical methods), which makes it difficult to assess real advantages in the NISQ era. The paper advocates for reporting standards, shared datasets and benchmarks, publication of code and circuits, and careful analysis of cost–accuracy–resources as necessary steps to mature quantum image processing and facilitate its practical adoption.

\section{Research networks}

During the QuantumX track, two key scientific collaboration initiatives were presented that play a strategic role in structuring and consolidating the Spanish and Ibero-American community around Quantum Computing and Quantum Software Engineering: the RIPAISC Network and the QSpain Network.

The Ibero-American Network for the Promotion of Quantum Software Engineering (RIPAISC) \cite{RIPAISC} is an international network that brings together researchers from Spain, Portugal, and Latin America with the aim of promoting scientific cooperation, training, and technology transfer in the field of quantum software. RIPAISC seeks to consolidate a cohesive Ibero-American community that addresses the emerging challenges of quantum development from an engineering, quality, and sustainability perspective, fostering collaboration between universities, technology centers, and companies. During the conference, the importance of this network as a platform for exchange for young researchers was highlighted, as well as its role in creating open repositories of tools, datasets, and teaching materials, and in organizing summer schools, thematic seminars, and joint projects.

QSpain is the Spanish association of quantum technology researchers \cite{QSPAIN}, acting as a national coordination hub between academic institutions, research centers, and companies working in the field of quantum computing, quantum communications, and post-quantum cryptography. QSpain's mission is to raise the profile of Spanish quantum research in the European context, encourage collaboration between groups, and promote Spain's coordinated participation in the European Union's Quantum Flagship initiatives and programs. Its presentation at QuantumX highlighted the growing interconnection between quantum hardware and quantum software, as well as the need to systematically incorporate software engineering principles into the development of reliable, scalable, and high-quality quantum algorithms, frameworks, and services.

The active participation of RIPAISC and QSpain underscored the importance of structured network collaboration as an essential strategy for advancing the scientific and technological maturity of Quantum Software Engineering, strengthening the Ibero-American and Spanish presence in major international research forums.

\section{Discussion}

The launch of the QuantumX track represented a significant milestone for the software engineering and quantum computing community in Spain. For the first time at SISTEDES, a dedicated space brought together both fields, showcasing recent scientific advances while also revealing research areas that remain largely open.

The conference was marked by an atmosphere of interdisciplinary collaboration, bringing together researchers in software engineering, computer architecture, artificial intelligence, databases, machine learning, and high-performance computing. The debates that followed the presentations highlighted the need to adapt classic software engineering methodologies (such as modeling, quality metrics, planning, orchestration, and testing) to the new paradigms of quantum programming and hybrid environments.

Among the most debated topics were efficient circuit planning in QPUs, the integration of quantum computing into hybrid cloud-edge-quantum architectures, and the need to define quality models and specific metrics for quantum software. The urgency of having more accessible high-level abstractions, frameworks, and quantum programming languages was also highlighted, allowing engineers to work without relying on low-level physical knowledge.

The networking sessions played a key role, allowing for the exchange of experiences between established groups and young researchers, as well as the identification of opportunities for collaboration in coordinated projects, doctoral programs, and European proposals.

Likewise, the presentation of the RIPAISC and QSpain networks reinforced the idea that community building and structured cooperation are essential to addressing the scientific and technological challenges of quantum software.

As an overall balance, the QuantumX track showed that the Spanish research ecosystem in quantum software engineering is mature, active, and expanding, with complementary lines of work and a growing international projection. The participating groups presented tangible results (such as articles, tools, projects, and prototypes) that reflect a shift towards more systematic and engineering-oriented approaches, positioning Spain as a relevant player in the evolving European quantum computing landscape.

\section{Conclusion}

The first edition of the QuantumX track, held within the framework of JISBD 2025, has clearly demonstrated both the scientific maturity and the growing dynamism of the Spanish research community working at the intersection of Quantum Computing and Software Engineering. The diversity of participating groups, the quality of the contributions, and the richness of the debates have shown that this emerging field is already evolving from theoretical reflection to experimental development, tooling, and practical engineering methodologies.

QuantumX has succeeded in positioning itself as a reference meeting point for researchers interested in transferring classical software engineering principles (such as quality assurance, testing, orchestration, and abstraction) to the quantum domain. The sessions have revealed a shared understanding that progress in quantum computing depends not only on advances in hardware but also on the creation of well-engineered, maintainable, and high-quality quantum software, supported by reproducible engineering practices. Taken together, the contributions discussed in this article provide a representative snapshot of current research directions and engineering concerns in Quantum Software Engineering, highlighting both achieved progress and remaining gaps.

Beyond the scientific presentations, the track fostered cross-institutional collaboration and community building, facilitated by the introduction of the RIPAISC and QSpain networks. These initiatives underscore the importance of sustained cooperation at both the national and Ibero-American levels to coordinate efforts, share resources, and promote the integration of Spain into European and global quantum programs.

The event has also served as a launch pad for future collaborations, with multiple groups identifying potential joint lines of work in topics such as hybrid architectures, quantum service governance, quality metrics, and quantum machine learning. The networking sessions and the diversity of research approaches (ranging from high-level programming abstractions to practical applications in AI and databases) illustrate the interdisciplinary potential of Quantum Software Engineering as a driver for innovation.

By bringing together these perspectives, the QuantumX track, and the analysis presented in this paper, contributes to clarifying the scope, challenges, and priorities of Quantum Software Engineering as an emerging discipline.

In light of its scientific impact, community engagement, and the enthusiasm generated among researchers and young scholars, QuantumX can be considered a resounding success. The experience validates the need to maintain and consolidate this track as a recurring space within SISTEDES, ensuring continuity and growth in upcoming editions. By providing an annual forum for discussion, collaboration, and dissemination, QuantumX is well-positioned to support the long-term consolidation of Quantum Software Engineering and to strengthen Spain’s role as an active contributor to the European and international quantum ecosystem.

%
%
%
\bibliographystyle{splncs04}
\bibliography{references}

\end{document}